# Design and implementation of an adaptive critic-based neuro-fuzzy controller on an unmanned bicycle

A. Shafiekhani[1], M. J. Mahjoob[1], M. Akraminia[1]

[1]Mechatronics Lab, School of Mechanical Engineering, College of Engineering, University of Tehran, Tehran, Iran.

Corresponding Author: M. J. Mahjoob, Associate Professor, Email Address: mmahjoob@ut.ac.ir , Tel: +98-21-66165538, Postal Address: Amirabad Street, College of Engineering, University of Tehran, Tehran, Iran.

**Abstract:**

Fuzzy critic-based learning forms a reinforcement learning method based on dynamic programming. In this paper, an adaptive critic-based neuro-fuzzy system is presented for an unmanned bicycle. The only information available for the critic agent is the system feedback which is interpreted as the last action performed by the controller in the previous state. The signal produced by the critic agent is used along with the error back propagation to tune (online) conclusion parts of the fuzzy inference rules of the adaptive controller. Simulations and experiments are conducted to evaluate the performance of the proposed controller. The results demonstrate superior performance of the developed controller in terms of improved transient response, robustness to model uncertainty and fast online learning.

**Keywords:** Adaptive control, Critic-Based control, Neuro-Fuzzy, Unmanned bicycle, Kalman filtering

## 1- Introduction

Bicycle is an interesting vehicle due to its help in human health and environmental issues as well as being an exciting sport tool for many decades. Balancing a bicycle by a human driver is possible; however, stabilizing an unmanned bicycle is very complicated. The first step in this control synthesis is understanding the bicycle dynamics which is a complex nonlinear system. Attempts have been made to investigate both the dynamics and control. Schwab et. al. presented several approaches such as pencil-and-paper, a numerical dynamics program, and a symbolic software to derive the linear motion equation of bicycle. This model has been a benchmark due to its accuracy [1]. Yavin [2] presented a nonlinear dynamic equation in which a simple structure of bicycle is used to develop equations of motion via the Lagrangian approach.

Different mechanisms have been applied to balance an unmanned bicycle which can be generally categorized in two groups, with or without stabilizer [3, 4]. Hwang et. al. applied two control strategies: first to control the bicycle center of gravity using an inverted pendulum and the second to control the steering angle [3] in which using two controller made the system more complicated. Chen employed an offline genetic algorithm to optimize the Fuzzy FIS (Fuzzy Inference System) membership functions in different forward velocities with handling of the steering angle [4].

The stabilizer based methods can also be classified into two groups: one using gyroscope (Control Moment Gyro (CMG) [5]) and the other inverted pendulum [6, 7]. A CMG has a spinning rotor and one or more motorized gimbals that change the axis of rotor's angular momentum. Changing the angular momentum creates gyroscopic torque and makes the bicycle stable.

Lam and Sin utilized a gyroscopic stabilizer and implemented a PD controller to make a typical bicycle stable [6]. High energy consumption and their further weight are the main drawbacks of these kinds of control schemes. Moreover, it can only be used to stablize the bicycle with no ability to track a specific path. Inverted pendulum has been also used to move the Center of Gravity (COG) and make the bicycle stable [8].

Unmanned bicycle can also be made stable by controlling the torque exerted on the steering handlebar with an actuator [9, 10]. Using this method, several control strategies have been used to stabilize the bicycle such as Fuzzy PID control [11], Fuzzy FIS [10] and Fuzzy sliding mode [3]. Considering the unstable nature of bicycle and the fact that the system is under-actuated, the controller desiging for this system becomes a highly challenging problem. Summing up the conclusion reached by the previous works, the control of a bicycle's roll angle and steering-handle angle are two of the most important issues in realizing a stable running motion. Whereas, in the human ride bicycle, both of these aspects are elegantly accomplished by body control and thus achieving stable bicycle motion.

In this study, a critic based neuro-fuzzy controller is employed to stabilize and control the bicycle. Critic gives rewards and/or punishments with respect to the states reached by the learner.

As critics constitute the less informative learning source, the learning methods using them represent very flexible tools [12]. These approaches, called reinforcement learning methods, consist of an active exploration of the state and action spaces to find what action to apply in each state [13]. This approach is also applied in other engineering areas; the authors in [14] employed the adaptive ciritic based controller to visually control a 7 DOF robot manipulator. This approach also used in [15] to tune steam generator water level.

We therefore focus on the design of this controller; a novel approach that can improve the transient response, robustness due to its model-free characteristic; the capability to adapt quickly with varying environment owing to learning ability. The main advantage of the proposed controller over previous fuzzy control approaches (e.g., neurofuzzy controller), is its online tuning characteristic by using a critic. That remarkably reduces the amount of computations used for parameter adaptation making it desirable for real time applications. This model free approach leads to a significant reduction in the computational burden as compared to model-based approaches, as well as existing learning approaches. The simplicity of the controller structure will make it attractive in industrial implementations where PD/PID type schemes are in common use [14]. In this reference, the computational load of this method and its convergence analysis by using direct method of Lyapunov method were studied.

The rest of paper is organized as follows. Next, a bicycle model is presented. In section 3, adaptive critic based controller is introduced. The approach is then tailored for an autonomous bicycle in section 4. Section 5 is dedicated to simulations, and section 6 presents the experimental results. Section 7 concludes the paper.

**2- System modelling**

The equations of motion of a bicycle forms the system model. Here we take a dynamic model consisting two DOF with fixed forward velocity [1]. Inputs to the model are steering handlebar angle and roll angle which are shown in Fig. 1 $\left(q = (\phi, \delta)^T\right)$. In this model, the velocity of bicycle is also considered fixed.

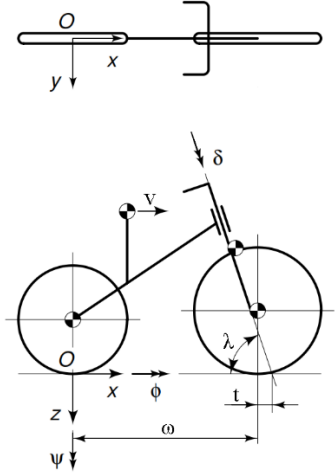

Fig 1. Bicycle model with demonstrating the coordinate system, the degrees of freedom and parameters[1]

In Fig. 1, $\delta$ shows the steering angle, $\phi$ is the roll angle, $\psi$ indicates the yaw angle and $v$ the forward velocity of bicycle, assumed constant here. The bicycle's equation of motion is:

$$M\ddot{q}^d + [C.v]\dot{q}^d + [K_1 + K_2 v^2]q^d = f^d \quad (1)$$

$$\dot{\psi} = \frac{v\delta + t\dot{\delta}}{\omega}\sin\lambda \quad (2)$$

$$\dot{x}_p = v\cos\psi$$
$$\dot{y}_p = v\sin\psi \quad (3)$$

where $M$ is the mass matrix whose elements are functions of components' mass and inertia. $C$ is the damping matrix, $K_1$ shows the velocity-independent elements of the stiffness matrix and $K_2$ shows the elements of the stiffness matrix to be multiplied by the square of the forward speed. The last element, f, is the external forces $\left(f = (f_\phi, f_\delta)^T\right)$ including steering force applied to the handlebar and lateral force which can be assumed as a disturbance. Considering the bicycle riding whether by human or an intelligent system, the steering force can be a torque applied by a hand or an actuator. Considering the rolling constraints, the yaw angle $\psi$ can be computed as a function of the steering angle, fixed velocity, and bicycle parameters [16]. $\lambda$ refers to the heading angle shown in Fig 1., it is called is called mechanical trail (i.e. the perpendicular distance that the front wheel contact point is behind the steering axis) and $\omega$ is the distance between centers of wheels. The above mentioned parameters are determined based on the size of bicycle used in our experiments as follows. The above mentioned parameters are determined based on the size of bicycle used in our experiments as follows [1]:

$$M = \begin{bmatrix} 1.43 & 0.18 \\ 0.18 & 0.08 \end{bmatrix} \quad (4)$$

$$C = \begin{bmatrix} 0 & 2.34 \\ -0.32 & 0.42 \end{bmatrix} \quad (5)$$

$$K1 = \begin{bmatrix} -32.53 & -4.3 \\ -4.3 & -1.6 \end{bmatrix} \quad (6)$$

$$K2 = \begin{bmatrix} 0 & 4.3 \\ 0 & 0.6 \end{bmatrix} \quad (7)$$

**3- Adaptive critic-based neurofuzzy controller**

**A. Neurofuzzy networks**

In this subsection, the principles of fuzzy system used here are introduced. An equivalent architecture is then formed that incorporates the fuzzy concept into an adaptive neural network. Generally, a fuzzy system consists of a fuzzification unit, a fuzzy rule base, an inference engine and a defuzzification unit. The fuzzy system can be viewed as performing a real (nonfuzzy) and nonlinear mapping from an input vector $X \in \mathbf{R}^n$, to an output vector $y = f(x) \in \mathbf{R}^m$ ($n$ and $m$ are dimensions of the input and output vectors, respectively). The interfaces between real world and fuzzy world are a fuzzifier and a defuzzifier; the former maps real inputs to their corresponding fuzzy sets and the latter performs in the opposite way to map the fuzzy sets of output variables to the corresponding real outputs. Two types of fuzzy systems are commonly used; Takagi-Sugeno-Kang (TSK) and fuzzy systems with fuzzifier and defuzzifier. In this work, we used the first type. The fuzzy rule base consists of fuzzy rules, which use linguistic If-Then statements to describing the relationship between inputs and outputs.

Consider a Multiple-Input Single-Output (MISO) fuzzy system consisting of N rules as follows:

$R_j$(jth rule): If ($x_1$ is $F_{j1}$) and ($x_2$ is $F_{j2}$) and ... and ($x_n$ is $F_{jn}$), Then $c_j = g_j(x)$.

where j=1,2, ... , N; $x_i$ (i=1,2, ... ,n) are the input variables of the fuzzy system, $F_{ji}$ is characterized by its corresponding membership function $\mu_{Fji}(x_i)$. $c_j$ which may generally be nonlinear, is the consequence of the jth rule and $g_j: R^n \rightarrow R$ is a general nonlinear or linear function. Each rule $R_j$, can be viewed as a fuzzy implication by the inference engine.

The antecedent fuzzy set (fuzzy Cartesian product) of each rule $F_1 \times F_2 \times ... \times F_n$, is quantified by the *t-norm* operator which may be defined as (8), the *min*-operator or the product operator

$$w_{F_1 \times ... \times F_n}(x_1,...,x_n) = \begin{cases} \min\left[\mu_{F_1}(x_1),...,\mu_{F_n}(x_n)\right] \\ or \\ \mu_{F_1}(x_1)...\mu_{F_n}(x_n) \end{cases} \quad (8)$$

The defuzzification is performed using (9), where $w_j$ is the firing strength of the antecedent part of the jth rule given by (10)

$$y = f(X) = \frac{\sum_{j=1}^{N} c_j w_j}{\sum_{j=1}^{N} w_j}, \qquad X = [x_1,...,x_n]^T \in R^n \quad (9)$$

$$w_j = \mu_{F_1}(x_1).\ ....\ \mu_{F_n}(x_n) \quad (10)$$

In this study, the consequent part of TSK fuzzy rules is given by

$$c_j = a_{0j} + \sum_{i=1}^{i=n} a_{ij} x_i \quad (11)$$

where $a_{0j}$ and $a_{ij}$ s are the coefficients that should be set at the design stage or tuned during the corresponding learning procedure.

Implementing a fuzzy inference system in the framework of an adaptive neural network leads to a five layer network in which each layer serves as one part of the equivalent fuzzy system. Fig. 2 shows a sample neurofuzzy system equivalent with a two-input and one-output TSK fuzzy inference system which has two linguistic labels for each input and therefore four rules in its rule base.

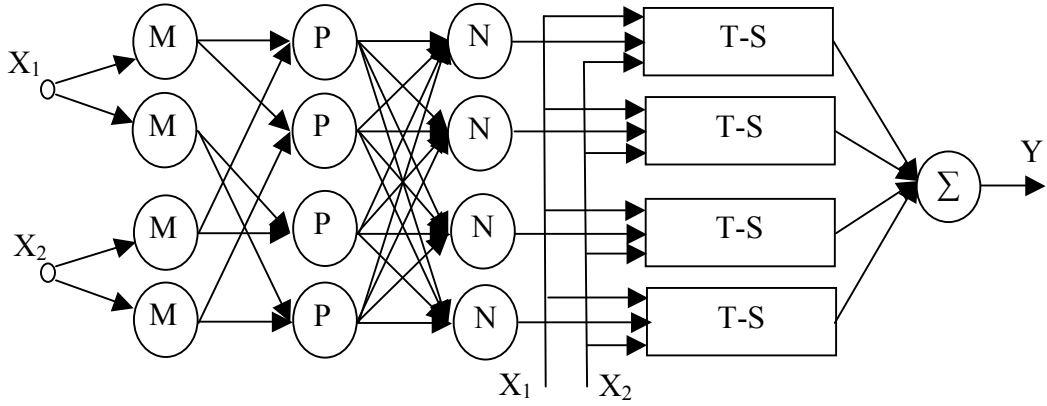

*Fig 2. A sample neurofuzzy structure equivalent with a MISO TSK fuzzy inference system*

The 1st layer nodes denoted by M, specifies the degree to which the given input satisfies the linguistic label; hence, they are fuzzy membership function for each input channel $\mu_{F_{ji}}(x_i)$. The 2nd layer nodes denoted by P, multiply the incoming signals and constitute the antecedent parts of fuzzy rules, (multiplication implies choosing the product-operator for *t-norm* operator). Each node in the 3rd layer specified by N, calculates the ratio of corresponding firing strength to the sum of all rules firing strengths; hence the term $w_j / \sum_{j=1}^{N} w_j$. The function of nodes in the 4th layer is performing a linear combination on inputs and adding a constant value, thus calculating the corresponding rule consequent part $c_j$. T-S labels on Fig. 2 refer to TSK rules. The coefficients of these linear combinations and that of constant value will be adapted during the learning stage. Finally, in the last layer, acting as the defuzzifier, the output is obtained according to (9).

**B. Controller structure**

Fig. 3 shows the proposed adaptive critic-based neurofuzzy control structure based on S. Russel and P. Norwig[17].

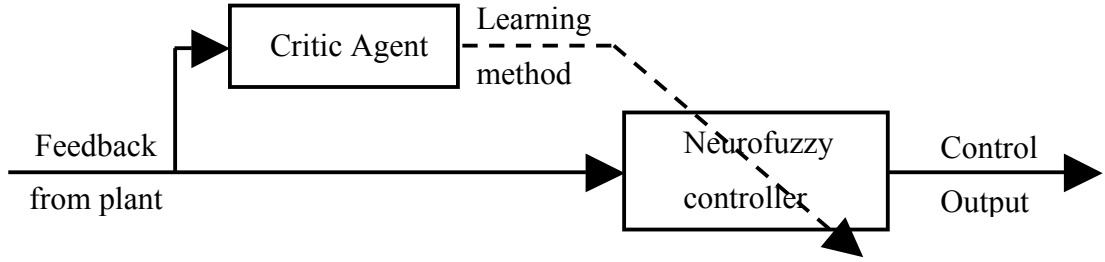

*Fig 3. Structure of adaptive critic-based neurofuzzy controller*

The critic agent evaluates the controller performance by assessment of the feedback from plant and generates appropriate reinforcement signal. The signal produced contributes collaboratively for updating parameters of the neurofuzzy controller. In classical reinforcement learning methods, the reinforcement signal accepts binary values, i.e., 1 for failure of the control action and 0 for the suitable performance. However, in modern approaches (e.g., [18]), the reinforcement signal is allowed to have real values e.g. in the range [-1,1], and the learning method is employed to adapt the tunable parameters of the controller in order to minimize this signal thus achieving 0 value indicating no need for further learning.

Let us define the cost function as

$$E = \frac{1}{2} r^2 \qquad (12)$$

where $r$ is the critic signal, the goal of the learning procedure is minimization of $E$, therefore the tunable parameters should be updated in the opposite direction of $\nabla E$ ($\nabla$ is the gradient operator). This can be stated as follows:

$$\Delta v \propto -\frac{\partial E}{\partial v} \qquad (13)$$

where $v$ is the tunable parameter of the neurofuzzy controller. Equation (13) is in fact the steepest decent law. It should be mentioned that other learning methods could also be used.

Applying the chain rule to calculate the partial derivative of (13),

$$\frac{\partial E}{\partial v} = \frac{\partial E}{\partial r} \frac{\partial r}{\partial u} \frac{\partial u}{\partial v} \qquad (14)$$

where $u$ is the control output. The aim of this study is to decrease the roll angle error to stabilize the bicycle. It is further required to reduce control effort to use a smaller force actuator as well as lower energy consumption for cost reduction. Hence we define the critic signal as a linear combination of error $e_\phi = \phi_{ref} - \phi$ and its rate $\Delta e_\phi(k) = e_\phi(k) - e_\phi(k-1)$,

$$r = k_1 e_\phi + k_2 \Delta e_\phi \tag{15}$$

where $k_1$ and $k_2$ are positive constants. Applying the chain rule and using (15),

$$\frac{\partial r}{\partial u} = \frac{\partial r}{\partial e_\phi} \times \frac{\partial e_\phi}{\partial y} \times \frac{\partial y}{\partial u} + \frac{\partial r}{\partial \Delta e_\phi} \times \frac{\partial \Delta e_\phi}{\partial y} \times \frac{\partial u}{\partial y} = -k_1 \frac{\partial y}{\partial u} \tag{16}$$

where $y$ is the system output. In (16), it is assumed that the sampling rate is small enough such that $\phi(k-1) \approx \phi(k)$. Substituting (16) in (13), we have

$$\frac{\partial E}{\partial v} = -k_1 r \frac{\partial y}{\partial u} \frac{\partial u}{\partial v} \tag{17}$$

In (17), the term $\frac{\partial y}{\partial u}$ is the gradient of the system and showing the long term variations of the plant output to the control signal. As in most cases, the system is designed such that this variation is a positive constant. The sign of this value, i.e., positive, is sufficient for the adaptation rule. Using (13) and (17), the learning rule for the tunable parameter will be:

$$\Delta v = \eta_1 \times k_1 \times r \times \frac{\partial u}{\partial v} \tag{18}$$

where $\eta_1 > 0$ is the learning rate which embeds the proportionality constant of (13) as well as the constant values of (17). Replace the constants product with a new constant $\eta$; therefore

$$\Delta v = \eta \times r \times \frac{\partial u}{\partial v}$$

For the neurofuzzy controller introduced in the previous subsection, the control signal has the following form (using (9) and (11)):

$$u = \frac{\sum_{j=1}^{N}(a_{0j} + \sum_{i=1}^{n} a_{ij} x_i) w_j}{\sum_{j=1}^{N} w_j} \tag{19}$$

Hence, according to (18) the updating rules for parameters of the neurofuzzy controller are:

$$\Delta a_{0j} = \eta \times r \times \frac{\partial u}{\partial a_{0j}} = \eta \times r \times \frac{w_j}{\sum_{j=1}^{N} w_j} \quad (20)$$

$$\Delta a_{ij} = \eta \times r \times \frac{\partial u}{\partial a_{ij}} = \eta \times r \times x_i \times \frac{w_j}{\sum_{j=1}^{N} w_j} \quad (21)$$

## 4. Controller design for unmanned bicycle

The structure of the adaptive critic-based neurofuzzy controller used for unmanned bicycle is shown in Fig. 4.

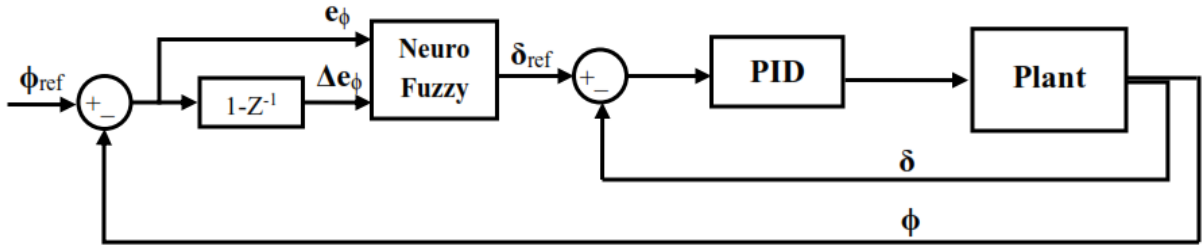

*Fig 4. Structure of the adaptive critic-based neurofuzzy controller for an unmanned bicycle*

In this figure, the plant incorporates the bicycle dynamics presented in section 2. The input to the plant is the steering torque $\tau_s$. The steering angle $\delta$ and the roll angle $\phi$ are the signals taken from the plant. The first signal, i.e. $\delta$, is applied within the internal loop to control the steering angle. The second signal (roll angle $\phi$), is used as feedback for the external loop in order to stabilize the bicycle using critic based neuro-fuzzy system. Critic based controller employs the roll angle error and its difference to produce an appropriate steering angle. The PID controller in the internal loop is then applied to generate accurate torque in order to make the error of steering angle zero.

The inputs of the neurofuzzy controller utilize three linguistic membership functions for each input, i.e., Negative (N), Zero (Z) and Positive (P), listed in table I and shown in Fig. 5. These membership functions may be extracted from system information, e.g. the range of variation of each input channel. The fuzzy rule base will then include nine rules.

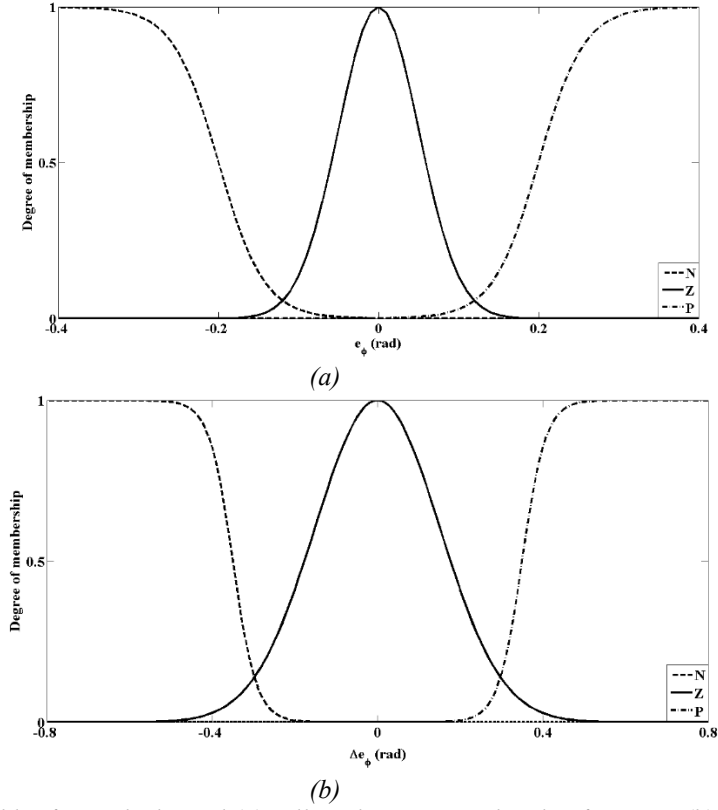

*(a)*

*(b)*

*Fig 5. Linguistic variables for each channel (a) roll angle error membership functions (b)difference of roll angle error membership functions*

*Table I. Linguistic variable functions for each input channel*

| Linguistic Variables<br><br>Input Channels | *Negative* | *Zero* | *Positive* |
|---|---|---|---|
| roll angle error (rad.) | $\dfrac{1}{1+e^{35(x+0.2)}}$ | $e^{\dfrac{-x^2}{2\times 0.05^2}}$ | $\dfrac{1}{1+e^{-35(x-0.2)}}$ |
| roll angle error difference (rad.) | $\dfrac{1}{1+e^{35(x+0.35)}}$ | $e^{\dfrac{-x^2}{2\times 0.15^2}}$ | $\dfrac{1}{1+e^{-35(x-0.35)}}$ |

It should be emphasized that the role of critic is to evaluate the situation; not accurate necessarily. Additionally, the critic can be a fuzzy system, a neural network or another system which has appropriate performance in evaluation of the system under control.

The proposed neurofuzzy controller has two inputs and employs nine rules in the fuzzy rule base whose consequent part are adaptive. Using (19) and (20), the parameters of the proposed adaptive controller will be updated as follows:

$$\Delta a_{0j} = \eta \times r \times \frac{w_j}{\sum_{j=1}^{N} w_j}, j = 1, 2, \ldots, 9 \quad (22)$$

$$\Delta a_{1j} = \eta \times r \times e_\phi \times \frac{w_j}{\sum_{j=1}^{N} w_j}, j = 1, 2, \ldots, 9 \quad (23)$$

$$\Delta a_{2j} = \eta \times r \times \Delta e_\phi \times \frac{w_j}{\sum_{j=1}^{N} w_j}, j = 1, 2, \ldots, 9 \quad (24)$$

where $\eta$ is the learning rate.

**5- Simulation results**

To compare the results, traditional fuzzy logic control [10] is applied to the model as well as our proposed method. The closed-loop system used here was shown in Fig. 4 . All simulations have been done in the MATLAB 13 and runned in the system with 2.4 GHz Core 2 Duo CPU and 2 GB RAM.

Deviations from equilibrium point and desired roll angle are investigated in this section. Appropriate values for $k_1, k_2$ and $\eta$ were obtained here by trial and error ($k_1 = 0.05, k_2 = 0.01$ and $\eta = 0.5$). The PID coefficients were also tuned to $K_P = 300, K_I = 100 \text{ and } K_D = 200$.

**Case 1.** In this case the forward velocity is 10 km/h, roll angle $\phi_{ref} = 0 \text{ and } \phi_{initial} = -15$. The controller task is to take the system back to its set point.

As the bicycle is unstable at low velocities, the main goal of this study is to obtain a controller which stabilizes the bicycle at low speeds. The performance of controllers for initial condition (initial roll angle is -15° here) are presented in Fig.6. Overshoot values of 1.1 and 5.9 deg. are observed for our proposed method and the Fuzzy FIS respectively (Fig. 6a). The settling time of our method is also 1 second while the Fuzzy FIS scores 3.1 seconds. The neuro-fuzzy controller has therefore offered better performance compared to the Fuzzy FIS approach. Fig. 6(b) also represents the output signal of the controllers which implies the faster response of our proposed method. Although our proposed method needs immediate action in the beginning and its undershoot is higher compared with Fuzzy FIS approach, it decreases the overshoot of the system response to the initial condition as well as the settling time.

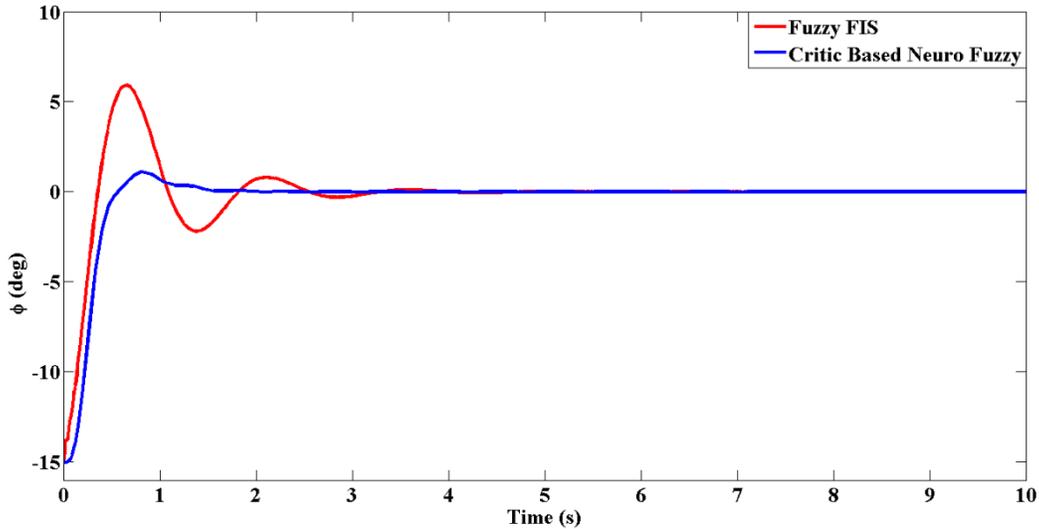

(a)

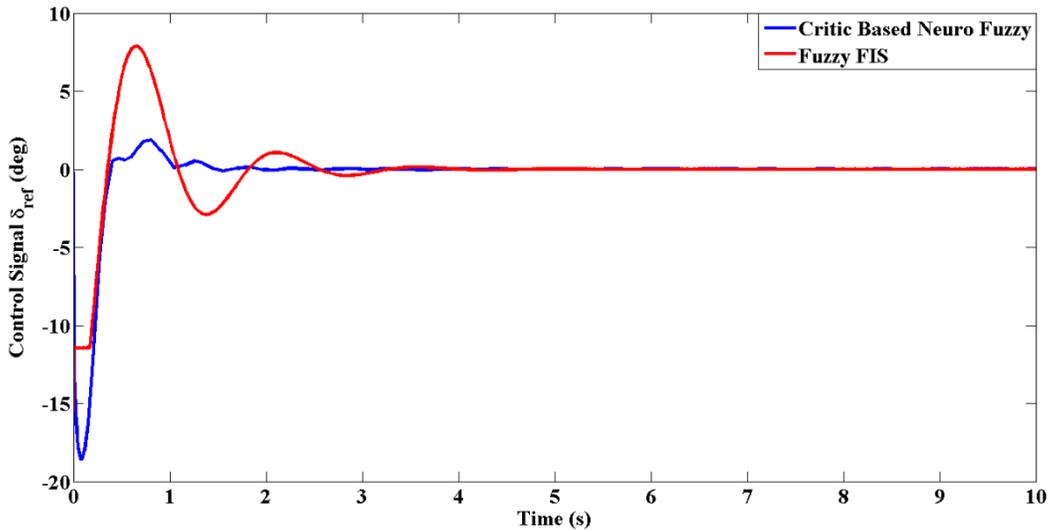

(b)

Fig 6. System output to $\phi_0 = -15$ for the case 1 of simulation (a) Roll Angle, (b) Control Signal

**Case 2.** Here, the reference roll angle $\phi_{ref}$ is considered to be a sinusoidal signal with frequency π/3 (rad/s) and amplitude $5$ . This can be a hard trajectory to be tracked both in simulation and experiments. The same closed-loop system is considered. Fig. 7(a) shows that the Fuzzy FIS deviates from the path at the Max. and Min. points while the Fuzzy critics based method closely follows the whole trajectory. While the bicycle roll angle follows a sine wave, the path travelled by mass center of the bicycle is then a sine curve in the X-Y plane as demonstrated in Fig. 7(b). This can later be used as a criterion to control the position of bicycle in the X-Y plane instead of roll angle of the bicycle.

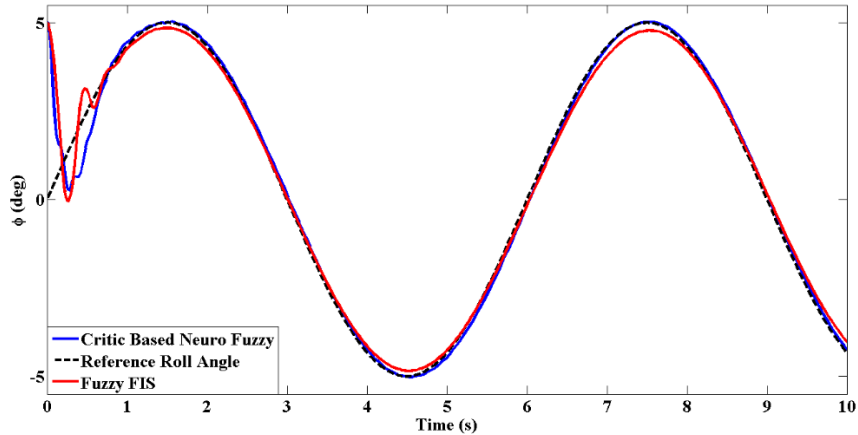

*(a)*

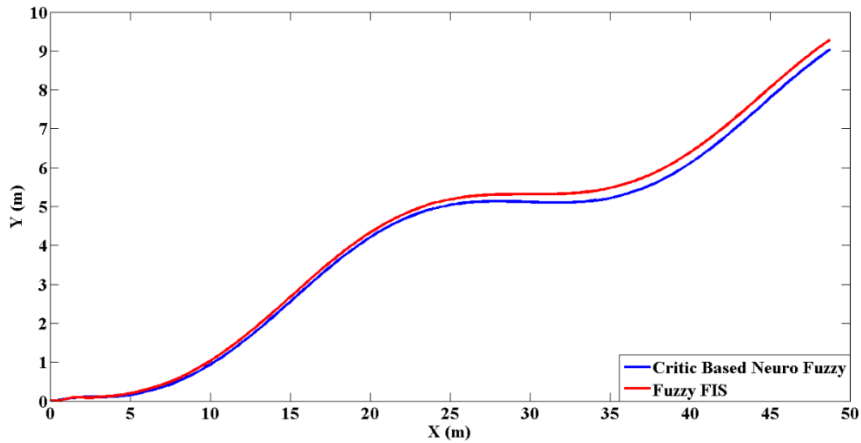

*(b)*

*Fig 7. Roll Angle Tracking performance of Fuzzy FIS and critic based neuro fuzzy for case 2 of simulation, (a) Roll Angle, (b) Trajectory of controlled bicycle with reference roll angle*

**Case 3.** To show the robustness of the citic based fuzzy system, in this case, a number of model parameters which can be changed in the real real world has been altered. Therefore, mass of the bicycle has increased by 20% and its velocity is set to 5 km/h. Then, the critic based fuzzy controller as well as the traditional fuzzy controller are applied to both pervious cases and the results are shown in the Fig 8.

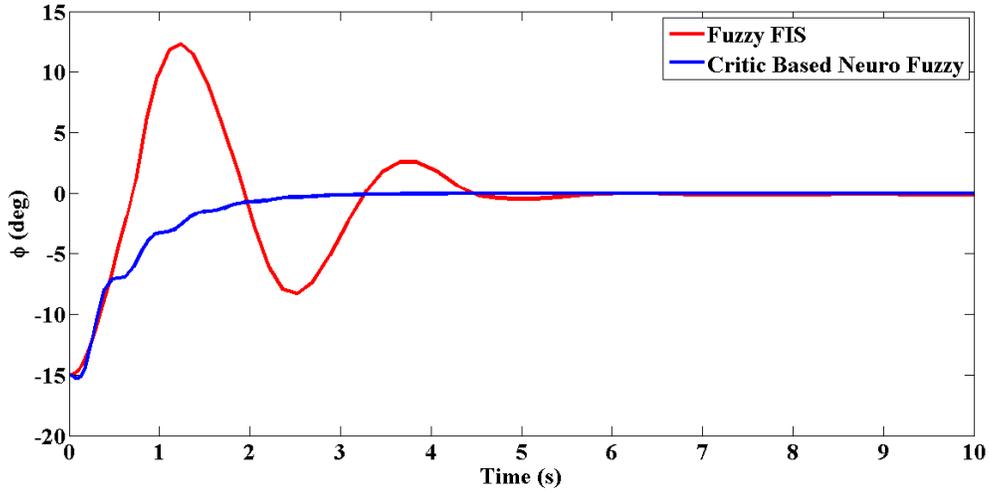

*(a)*

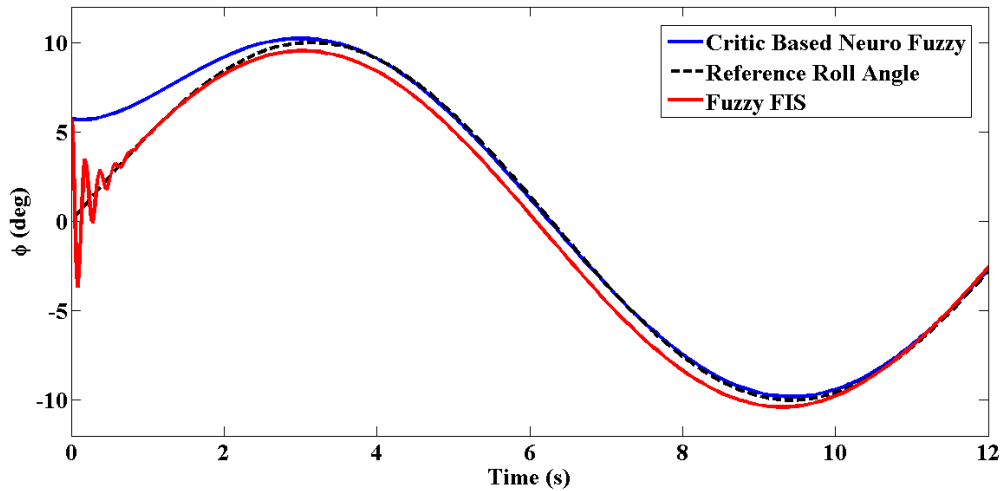

*(b)*
*Fig 8. Roll Angle Tracking performance of Fuzzy FIS and critic based neuro fuzzy for case 2 of simulation, (a) Roll Angle, (b) Trajectory of controlled bicycle with reference roll angle*

Overshoot value of 12.3 deg. is observed Fuzzy FIS while our proposed method offers no overshot (Fig. 8a). The settling time of our method is also 2.9 second while the Fuzzy FIS scores 4.2 seconds. In addition, Fig. 8b shows the trackong error where our proposed method has converged to the reference signal after 3.2 seconds. In contrast, Fuzzy FIS approach gives an oscillation about the reference signal at the intialization as well as and offset to the sinusoidal reference signal. Consequently, our proposed method provides improved performance in this case compared to the traditional Fuzzy approach. This better performance is due to the ability of our method in the online tuning of its adaptive parameters such that the cost function approaches to the minimum value.

## 6- Experimental test setup

Experiments are designed and conducted here to investigate the capability of our proposed method in practice. The main part of the test setup is a bicycle equipped with an IMU (type MPU6050) which measures the roll angle of bicycle (φ), and a 200 pulse incremental encoder to measure the steering rotation (δ) (Fig. 9 & 10). A microcontroller board (Arduino type UNO) has been used. A geared DC motor produced appropriate torques to turn the bike steering fork. The motor based on our test had 2.5 N.m stall torque and reached a no-load speed of 130 rpm at 24v. The motor position control was made by a PID controller whose coefficients were tuned experimentally $(K_P = 40, K_I = 0.001 \ and \ K_D = 1)$. A treadmill was used to implement the driving in the lab and run the bicycle with desired velocities according to our test conditions.

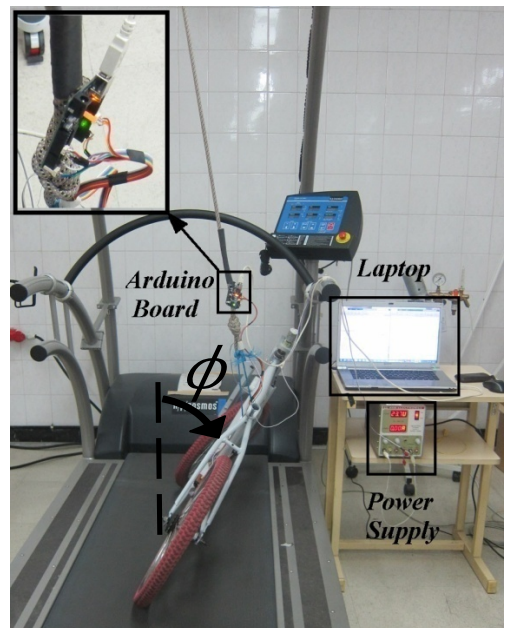

*Fig. 9. Experimental test set-up*

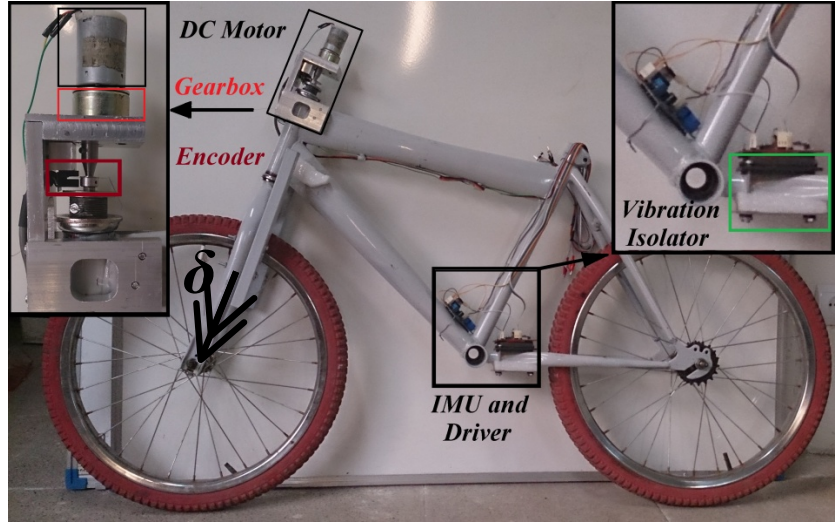

*Fig 10. Prototype equipped bicycle with sensors and actuator*

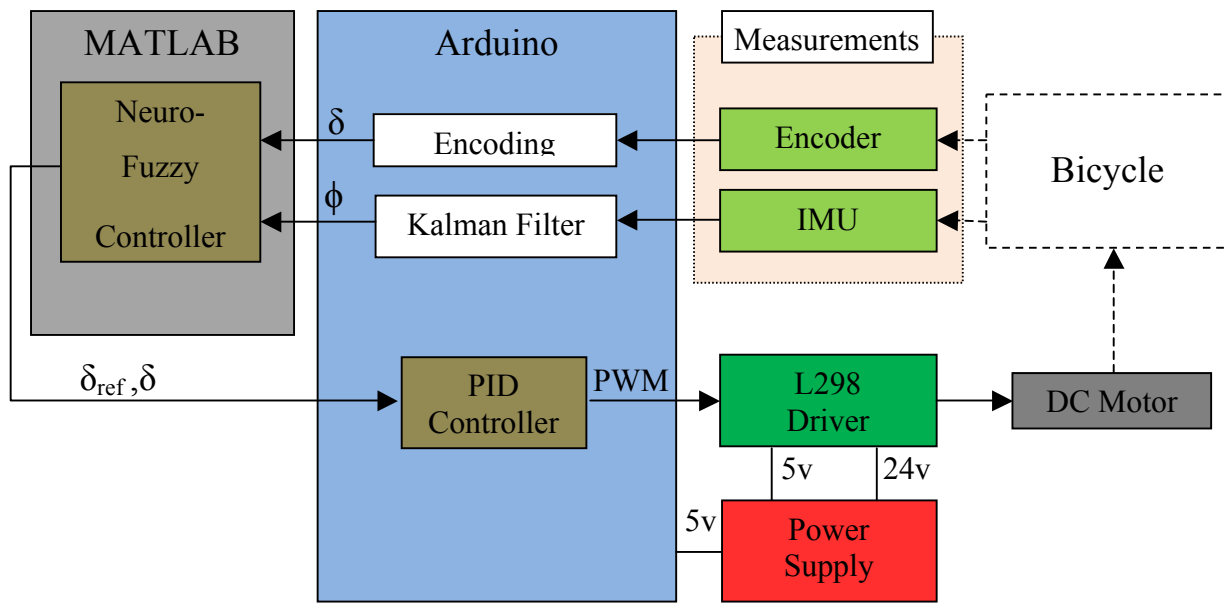

*Fig 11. Control and measurement loop*

### A. Signal conditioning

In order to filter out the noise from MPU6050 data, 260 Hz and 256 Hz low pass filters are applied to the accelerometer and gyro data correspondingly. The accelerometer is generally noisy while the gyro drifts over time (the gyro data can be used in a shorter time while the accelerometer data may be utilized much longer). To treat this, a Kalman filter is designed/implemented in which a combination of accelerometer and gyro data are employed to obtain a better roll estimation by removing the unwanted residual noise (Fig. 12 and 13).

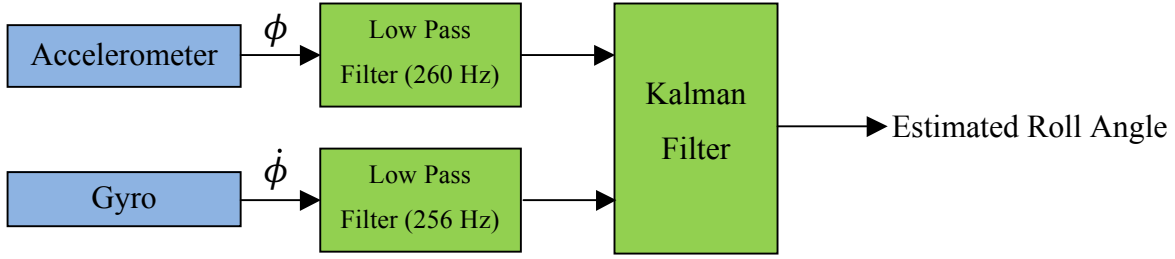

Fig 12. Estimation of roll angle from IMU data

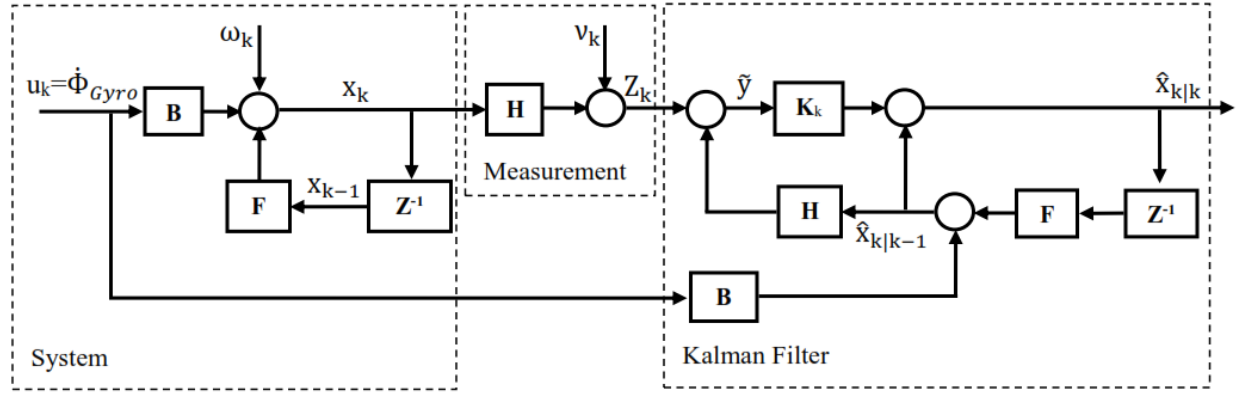

Fig 13. Block diagram of system, measurement, and discrete Kalman filter

The system equations are

$$x_k = Fx_{k-1} + Bu_k + \omega_k \tag{25}$$

$$z_k = Hx_k + v_k \tag{26}$$

in which the state vector is $x_k = \begin{bmatrix} \phi(k) \\ \phi_b(k) \end{bmatrix}$ where $\phi(k)$ is the roll angle measured by IMU and $\phi_b(k)$ shows the bias indicating gyro drift at the time $k$. The input to the system $u_k$ is $\phi_{Gyro}(k)$; adopted originally from accelerometer. Other matrices are $F = \begin{bmatrix} 1 & -\Delta t \\ 0 & 1 \end{bmatrix}$, $B = \begin{bmatrix} \Delta t \\ 0 \end{bmatrix}$ and $H = \begin{bmatrix} 1 & 0 \end{bmatrix}$; $\Delta t$ shows the time step. The output $z_k$ of this system is $\phi(k)$. $\omega_k$ indicates the process noise which is Gaussian with zero mean and covariance $Q$ at the time $k$; i.e.

$$\omega_k \sim N(0, Q_k) \tag{27}$$

$$Q_k = \begin{bmatrix} Q_\phi & 0 \\ o & Q_{\phi_b} \end{bmatrix} \Delta t \tag{28}$$

Similar to the process noise, measurement noise $v_k$ is represented by a Gaussian noise with zero mean and variance $R$ at time $k$,

$$v_k \sim N(0,R) \text{ and } R = \mathrm{var}(v_k) \tag{29}$$

The final roll angle is then estimated by the Kalman filter as follows [19]:

$$\hat{x}_{k|k} = \hat{x}_{k|k-1} + K_k y \tag{30}$$

$$\hat{x}_{k|k-1} = F\hat{x}_{k-1|k-1} + B\phi_k \tag{31}$$

$$K_k = P_{k|k-1} H^T S_k^{-1} \tag{32}$$

$$P_{k|k-1} = FP_{k-1|k-1}F^T + Q_k \tag{33}$$

$$S_k = HP_{k|k-1}H^T + R \tag{34}$$

$$y = z_k - H\hat{x}_{k|k-1} \tag{35}$$

$$P_{k|k} = (I - K_k H)P_{k|k-1} \tag{36}$$

$\hat{x}_{k|k}$, called posteriori state, is the estimation of $x_k$ at step $k$ based on the measurement at the same time step. $\hat{x}_{k|k-1}$ or priori state, is the estimate of the state at the current time $k$ based on the previous state of the system, $P_{k|k-1}$ shows the a priori error covariance matrix, $P_{k|k}$ demonstrates posteriori error covariance matrix and $S_k$ illustrates the innovation covariance. In this study, the covariance matrix $Q$ and variance $R$ are $Q_k = \begin{bmatrix} Q_\phi & 0 \\ 0 & Q_\phi \end{bmatrix} = \begin{bmatrix} 0.001 & 0 \\ 0 & 0.003 \end{bmatrix}$ and $R = \mathrm{var}(v) = 0.03$.

Typical output of the Kalman filter along with the original accelerometer and gyro signals are shown in the Fig. 14.

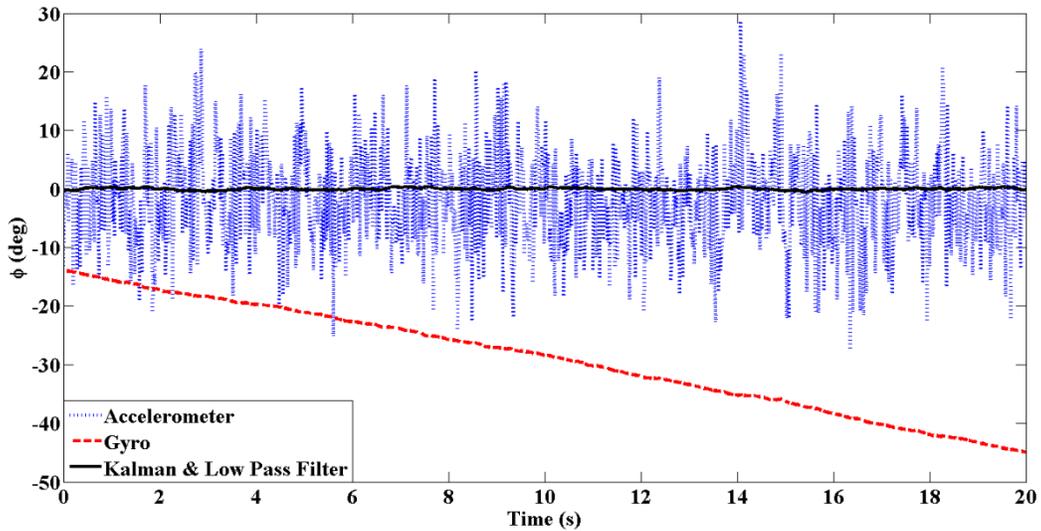

Fig 14. Accelerometer, Gyro and their filtered data with Kalman and low pass filter

**B. Results**

Experiments are now performed to verify what we developed here. The overall system is shown schematically in Fig.15.

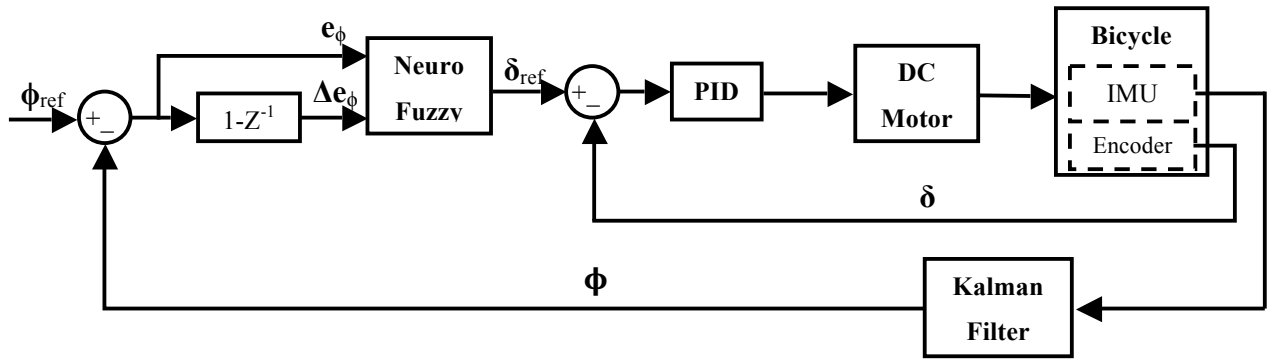

Fig 15. Block diagram of the implemented control loop

This block diagram is similar to Fig. 4 which was already used in the simulations. In this structure, DC motor is added and the outputs of IMU sensor are filtered via the Kalman filter. The adjustments were made in the test setup compared to the blocks used in the simulation. The PID were experimentally tuned to $k_p = 40, k_I = 1$ and $k_D = 0.01$. The cost function coefficients $k_1, k_2$ and learning rate $\eta$ were also set to 0.4, 0.3 and 1.0 correspondingly. The control performance in stabilizing the system as well as tracking the desired roll angle trajectory was investigated in the following cases:

**Case 1.** A forward velocity of 10 km/h and an initial roll angle of -18° are applied to the bicycle. The controllers are expected to drive the roll angle to its reference set point ($\phi_{ref} = 0$).

Fig. 16(a) demonstrates the roll angle (degree) versus time (second). Solid-blue line indicates the result of Neuro Fuzzy controller while the dashed-red curve shows the FIS result. The roll angle overshoot for Fuzzy FIS and our proposed method are 12.2 and 5.8 deg., respectively. Less overshoot was therefore achieved by the neuro-fuzzy method as well as smaller amount of oscillation to reach the set point. It is useful to compare the results of simulation and experiment in this case (figures 6 and 16). From these figures one can observe that, unlike the experiment, fast convergence rate was obtained by using the proposed method in the simulation. Delays in the response of the DC motor due to its inertia can be a main reason for this oscillatory behavior. Fig. 16(b), which demonstrates the control and steering angle signals, also shows that the motor cannot closely follow the control signal in both methods.

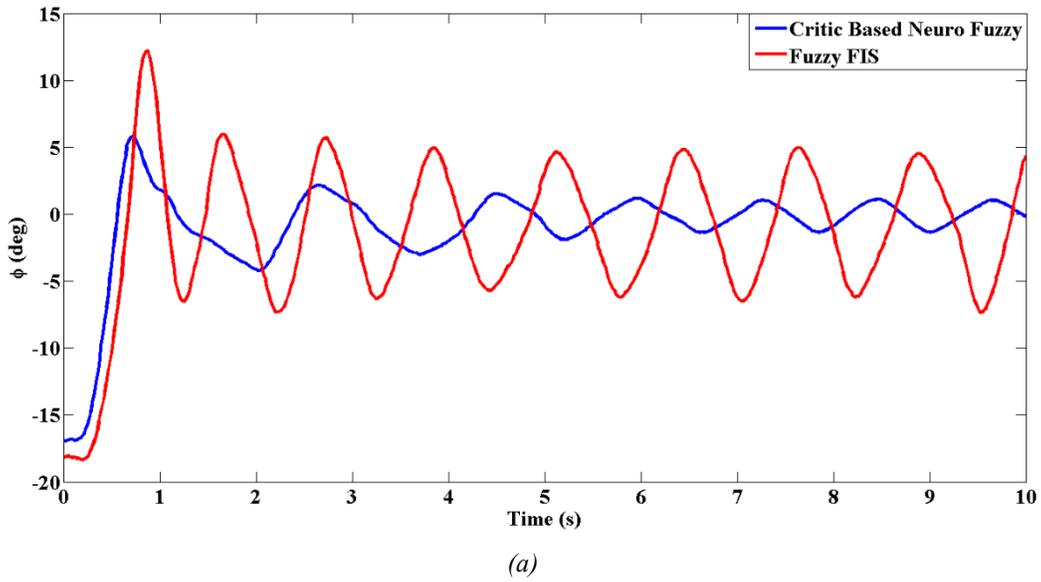

*(a)*

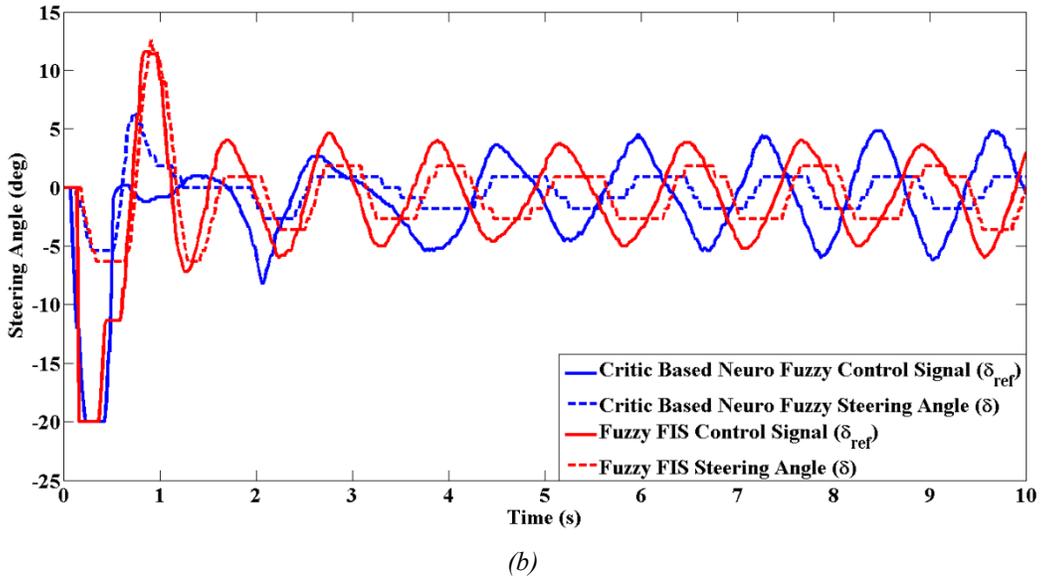

*(b)*

Fig 16. System output to $\phi_0 = -18$ in case 1 of experiment, (a) Roll angle, (b) Steering angle

**Case 2.** The forward velocity is the same as in previous case; however, a sinusoidal roll angle trajectory is employed with frequency $\pi/3$ radian/s and amplitude 5 deg. The results of two methods are shown in the Fig. 17. The same line notation is used here as in the previous case. Fig. 17 shows a better tracking performance for neuro-fuzzy method compared to the fuzzy FIS approach. It also shows that the tracking ability of our proposed method is improved step by step due to its learning ability.

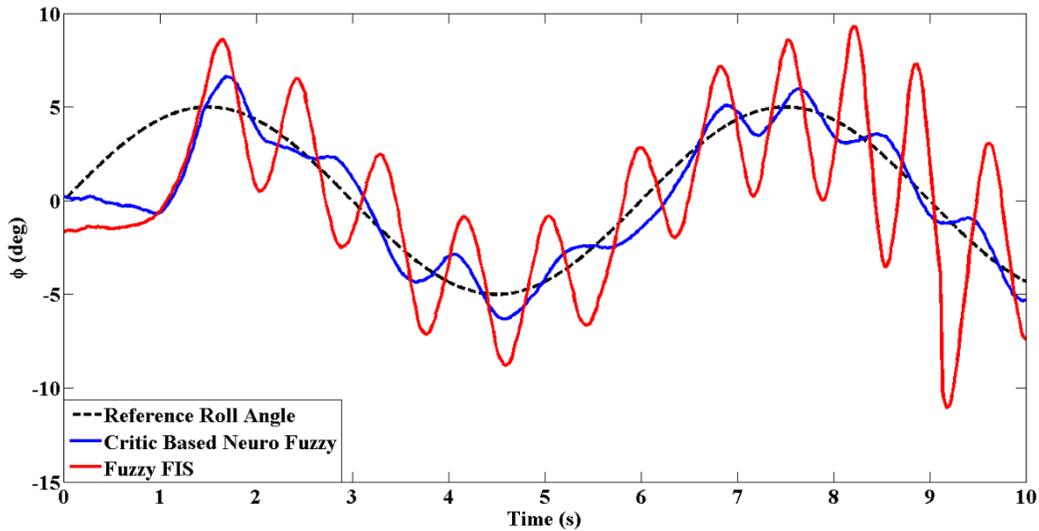

*Fig 17. Experimental results of Roll Angle Tracking performance of Fuzzy FIS and critic based neuro fuzzy*

## 7- Conclusion

An adaptive critic-based neuro-fuzzy controller was introduced to balance an unmanned bicycle. The signal produced by the critic agent along with an error back-propagation algorithm were used to adjust the conclusion parts of the fuzzy inference rules of the adaptive controller. Simulations were then performed to compare the results of the Fuzzy FIS system with our proposed method. A test setup was also designed and the control methods were evaluated experimentally. Due to noisy outputs of the IMU sensor, a Kalman filter was applied to obtain better IMU signals in the experiments. Both simulation and experimental results showed some superior performance of the proposed control method in terms of its simplicity, improved transient response, robustness and fast online learning.